\documentclass[prd,final,aps]{revtex4}
\usepackage{graphicx}
\usepackage{amsmath}
\usepackage{amssymb}
\usepackage{hyperref}
\usepackage{bm}

\def\g{\gamma}

\def\p{\pi}

\def\D{\Delta}

\def\L{\Lambda}
\def\O{\Omega}

\begin{document}
\preprint{BROWN - HET- 1714}
\title{Effect of the Cosmological Constant on Halo Size}
\author{Ekapob Kulchoakrungsun, Adrian Lam, David A. Lowe} 
\affiliation{Department of Physics, Brown University, Providence, RI 02912, USA}

\begin{abstract}
In this work, we consider the effect of the cosmological constant on galactic halo size. As a model, we study the general relativistic derivation of orbits in the Schwarzschild-de Sitter metric. We find that there exists a length scale $r_\Lambda$ corresponding to a maximum
size of a circular orbit of a test mass in a gravitationally bound system, which is the geometric mean
of the cosmological horizon size squared, and the Schwarzschild radius. This agrees well with the size of a galactic halo when the effects of dark matter are included. The size of larger structures such as galactic clusters and superclusters are also well-approximated by this scale. This model provides a simplified approach to computing the size of such structures without the usual detailed dynamical models.  Some of the more detailed approaches that appear in the literature are reviewed, and we find the length scales agree to within a factor of order one. Finally, we note the length scale associated with the effects of MOND or Verlinde's emergent gravity, which offer explanations of the flattening of galaxy rotation curves without invoking dark matter,
may be expressed as the geometric mean of the cosmological horizon size and the Schwarzschild
radius, which is typically 100 times smaller than $r_\Lambda$.

\end{abstract}
\maketitle
\newpage

%%%%%%%%%%%%%%%%%%%%%%%%%%%%%%%%%%%%%%%%%%%%%

\section{Introduction}
Price and Romano \cite{Price:2005iv} asked ``In an expanding universe, what doesn't expand?" In their paper, they considered the effect of cosmological expansion on a classical atom bound by electrical attraction, and analyzed its stability. In this paper, we then ask similar questions, but on the galactic scale: How does the cosmological expansion affect the rotation curve of the galaxy, and at what scale does this effect dominate? To answer these questions, we present a general relativistic calculation of the angular velocity for a circular orbit of a test mass in a Schwarzschild-de Sitter metric. We find that there exists a dynamical length scale corresponding to the maximum size of a circular orbit of a test mass in a gravitationally bound system. We confirm this length scale is comparable to the dark matter halo size definition commonly used in the literature \cite{White:2000jv,Jenkins:2000bv}. We then compare our length scale to the ones considered in MOND and Verlinde's emergent gravity.

The paper is organized as follows. Section \ref{sec:velocity calculation} gives the derivation for the angular velocity as a function of radius in Schwarzschild-de Sitter spacetime. We obtain a length scale $r_\L$ characterizing the size of the gravitationally bound system. In section \ref{sec:Halo size}, we discuss dark matter halo size as derived from simulation involving collapse and virialization, and show there is a simple relation with the scale $r_\L$. Section \ref{sec:MOND} presents the summary of MOND and Verlinde's emergent gravity, and the corresponding length scales where new physics appears, which are parametrically smaller than $r_\L$, though curiously built out of a different geometric mean of the cosmological horizon size and the Schwarzschild radius.
%%%%%%%%%%%%%%%%%%%%%%%%%%%%%%%%%%%%%%%%%%%%%%%%%%%%%%%%%%%%%%%%%%%%%%%
\section{Schwarzschild-De Sitter Orbits}\label{sec:velocity calculation}

We present the general relativistic derivation of the angular velocity for a circular orbit of a test mass in the Schwarzschild-de Sitter metric 
\begin{equation*}
    ds^2 = -\left(1-\frac{2M}{r} - \frac{1}{3}\Lambda r^2\right) dt^2 + \left(1-\frac{2M}{r} - \frac{1}{3}\Lambda r^2\right)^{-1} dr^2 + r^2(d\theta^2+\sin^2\theta d\phi^2),
\end{equation*}
where $M$ is the mass enclosed by the orbit, and $\Lambda>0$ is the cosmological constant. Note that this form of the metric will be valid outside general spherically symmetric mass distributions by Birkhoff's theorem.

For orbits, spherical symmetry means the motion is confined to a plane, which we choose to be the equatorial plane given by $\theta = \pi/2$.  Since the metric $g_{\mu\nu}$ is independent of $t$ and $\phi$, we identify the two Killing vectors as
\begin{eqnarray*}
  K^\mu &=& \partial_t^\mu = (1,0,0,0),\\
  R^\mu &=& \partial_\phi^\mu = (0,0,0,1),
\end{eqnarray*}
and the two corresponding conserved quantities are given by
\begin{eqnarray*}
  E &=& -K_\mu\frac{dx}{d\tau}^\mu = \left(1-\frac{2M}{r} - \frac{\Lambda r^2}{3}\right)\frac{dt}{d\tau},\\
  L &=& R_\mu\frac{dx}{d\tau}^\mu = r^2 \frac{d\phi}{d\tau}.\label{eq:def L}
\end{eqnarray*}
The invariant spacetime interval
\begin{equation*}
  g_{\mu\nu}\frac{dx}{d\tau}^\mu\frac{dx}{d\tau}^\nu = -1
\end{equation*}
can then be expressed in terms of these two conserved quantities as 
\begin{equation}\label{eq:EOM}
 \left(\frac{dr}{d\tau}\right)^2 = E^2 - \left(1-\frac{2M}{r}-\frac{\Lambda r^2}{3}\right)\left(1+\frac{L^2}{r^2}\right).
\end{equation}
It is then convenient to identify the effective potential as
\begin{equation}\label{eq:V}
  V_\mathit{eff} \equiv \left(1-\frac{2M}{r}-\frac{\Lambda r^2}{3}\right)\left(1+\frac{L^2}{r^2}\right).
\end{equation}

A circular orbit is possible only at a minimum or a maximum of $V_\mathit{eff}$, and we must have $dr/d\tau = 0$. We differentiate \eqref{eq:V},
\begin{equation}
  \frac{dV_\mathit{eff}}{dr} = 0,
\end{equation}
which gives
\begin{equation}\label{eq:L}
  \frac{L^2}{r^2} = \left(\frac{M}{r} - \frac{\Lambda r^2}{3} \right)\left(1- \frac{3M}{r} \right)^{-1} 
\end{equation}
Using \eqref{eq:EOM} the energy is
\begin{equation}\label{eq:E}
  E^2 = \left(1-\frac{2M}{r}-\frac{\Lambda r^2}{3}\right)\left(1+\frac{L^2}{r^2}\right) = \left(1-\frac{2M}{r}-\frac{\Lambda r^2}{3}\right)^2 \left(\frac{dt}{d\tau}\right)^2.
\end{equation}
Substituting $L$ from Eq.(\ref{eq:L}) into Eq.(\ref{eq:E}) gives
\begin{equation}
  \left(\frac{dt}{d\tau}\right)^2 = \left(1-\frac{3M}{r}\right)^{-1}.
\end{equation}
We also have from Eq.(\ref{eq:def L}) and Eq.(\ref{eq:L}) 
\begin{equation}
  r\frac{d\phi}{d\tau} = \left[\left(\frac{M}{r} - \frac{\Lambda r^2}{3} \right)\left(1- \frac{3M}{r} \right)^{-1}\right]^{1/2} .
\end{equation}
We finally obtain the angular velocity
\begin{equation}\label{eq:angular velocity}
  v_\phi = r\frac{d\phi}{dt} = r\frac{d\phi/d\tau}{dt/d\tau} = \sqrt{\frac{M}{r} - \frac{\Lambda r^2}{3} }
\end{equation}
which implies that there is a maximum radius where the angular velocity becomes zero.
Thus we conclude there is no circular orbit beyond the radius $r_\Lambda$ given by 
\begin{equation}\label{eq:r_lambda}
	r_\Lambda = \left(\frac{3GM}{\Lambda c^2}\right)^{1/3},
\end{equation}
where the factors of $G$ and $c$ are restored. This is simply proportional to the geometric mean of the Schwarzschild radius and the length of the cosmological horizon squared.

\begin{figure}[t]
\includegraphics[width=15cm]{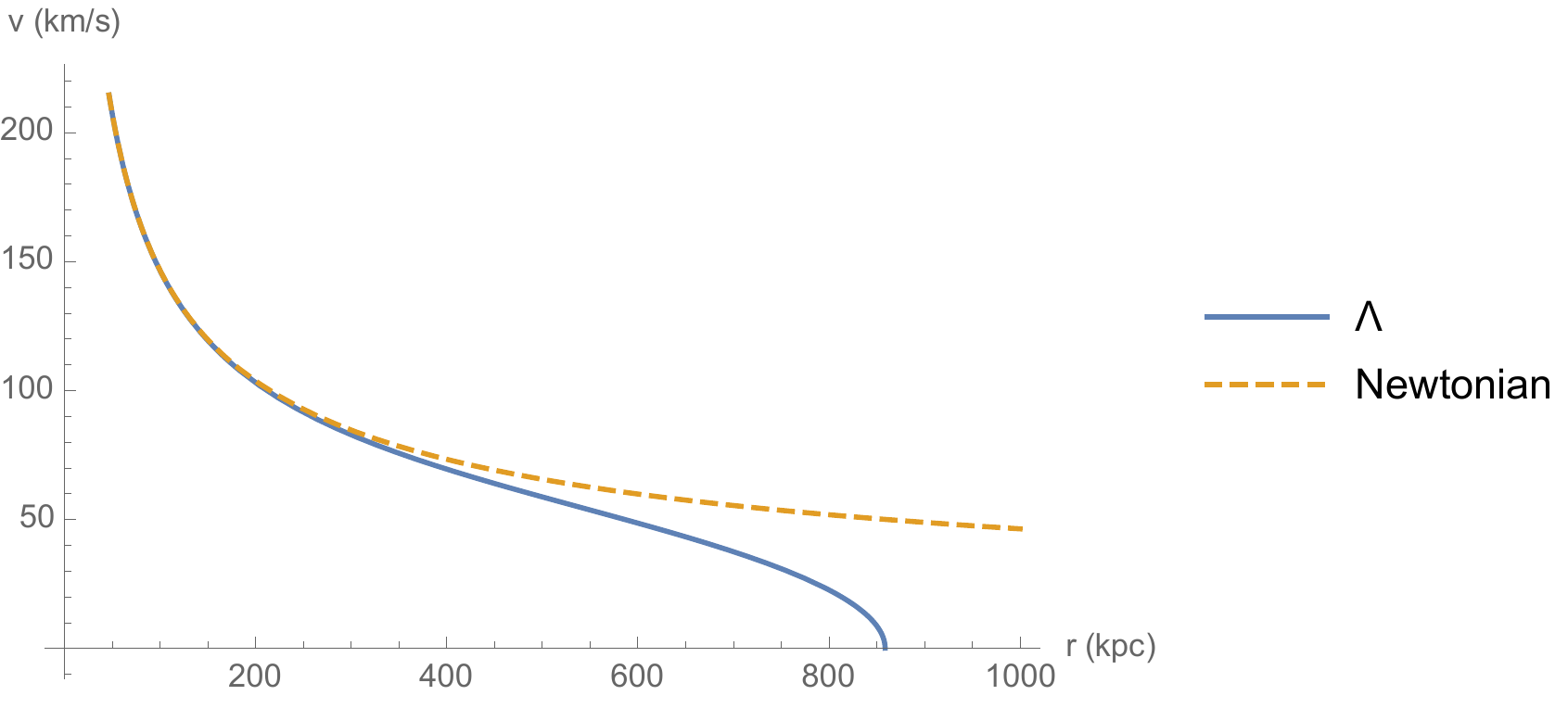}
\caption{Angular velocity as a function of radius $r$ in kpc, for the Triangulum Galaxy (M33) with the total mass taken as $5\times10^{11}M_{\odot}$ \cite{m33}.}\label{fig:velocity}
\end{figure}

The resulting angular velocity, in contrast to the Newtonian velocity, is plotted in Fig.\ref{fig:velocity}. We interpret this maximum radius, $r_\Lambda$, as the maximum size of the gravitationally bound object. This length scale should be comparable to the size of a dark matter halo. We can check this by comparing our length scale to the halo size definitions used in the literature \cite{White:2000jv}. A similar simplified strategy to calculating halo size as presented here has been previously considered in \cite{Bhattacharya:2010gm} in a more general context, and the emergence of the length scale (\ref{eq:r_lambda}) was not noted. 

Upon completion of this paper, the authors came across related work in \cite{Souriau, Stuchlik1, BalagueraAntolinez:2005wg, Mizony, Stuchlik2}. An early appearance of the length scale (\ref{eq:r_lambda}) can be found in \cite{Souriau}, which was also derived in \cite{Stuchlik1} and \cite{BalagueraAntolinez:2005wg} through similar considerations. In \cite{BalagueraAntolinez:2005wg}, the length scale (\ref{eq:r_lambda}) was numerically compared to the order of magnitude of typical sizes of astrophysical multi-body objects, namely globular clusters, galaxies, galaxy clusters, and superclusters, by taking $M$ as the average mass of the members comprising the object. The importance of accounting for dark matter in the estimate of $M$ was not noted. Considerations of the possible significance of cosmological effects on rotation curves and dark matter estimations were mentioned in \cite{Mizony} and \cite{Stuchlik2}.

In the present work we take $M$ to be the total mass of the structure in (\ref{eq:r_lambda}), which in general includes dark matter. The length $r_\L$ is identified as the maximum radius at which orbiting objects like stars, which we can safely treat as test particles, are still gravitationally bound.   With this in mind, Table \ref{table:struc} shows the length scales predicted by $r_\L$ for several astrophysical structures. The second row pertains to the visible components of the galaxy which is dominated by the stellar disk and gas components, while the third row refers to the entire galaxy, which is dominated by its dark matter halo.  Correspondingly, the masses of galaxy clusters and superclusters are also dominated by dark matter. It is clear from Table \ref{table:struc} that $r_\L$ gives an accurate measure of size for scales of galactic dark matter halos and beyond, but is of little relevance for scales below that.  The inclusion of dark matter is crucial here for obtaining the structure size.

In the next section, we compare the length scale (\ref{eq:r_lambda}) with halo size derived from the spherical top-hat collapse model, an analytic model commonly used in the literature to predict the endstate of virialized halos.  As we will see, $r_\L$ and the virial radius of halos have the same analytic form up to a dimensionless constant of order 1.

\begin{table}
  \begin{tabular}{ | l | l | c | c | c |}
    \hline
    \textbf {Structure} & \textbf{Example} & \textbf{Total Mass} ($M_{\odot}$) & \textbf{Actual Size \textit{r}} (pc)& ~~~~\boldsymbol{$r_{\Lambda}$} (pc)~~~~ \\ 
    \hline
    Globular Cluster & Messier 30 \cite{messier,glob2}& $1.6\times10^5$ & 14 & $6.0\times10^3$ \\ \hline
    Galaxy (visible matter) & Triangulum Galaxy (M33) \cite{messier,m33} & $1\times10^{10}$  & $8.3\times10^3$ & $2.4 \times 10^5$ \\ \hline
    Galaxy (total matter) & Triangulum Galaxy (M33) \cite{m33} & $5 \times10^{11}$  & $1.8\times10^5$ & $8.7\times10^5$ \\ \hline
    Galaxy Cluster & Virgo Cluster \cite{Virgo}	& $8.0\times10^{14}$ & $7.2\times10^6$ & $1.0\times10^{7}$ \\ \hline
    Supercluster & Laniakea Supercluster \cite{Laniakea}&$1\times10^{17}$&$ 8\times10^{7}$  & $5.1\times10^7$ \\ \hline
  \end{tabular}
  \caption{Comparison of actual size and size predicted by $r_{\Lambda}$ for various structures, taking $M$ as the total mass of the structure. The third row gives the total mass and extent of the entire galaxy, which is largely dominated by dark matter, and so describes the dark matter halo.  Likewise, the masses for galaxy cluster and supercluster are dominated by dark matter.  We have included superclusters for completeness, but note that most superclusters are not gravitationally bound; they are not virialized and will disperse in the future \citep{Chon}.}
  \label{table:struc}
  \end{table}

%%%%%%%%%%%%%%%%%%%%%%%%%%%%%%%%%%%%%%%%%%%%%%%%%%%%%%%%%%%%%%%%%%%%%%%
\section{Halo Size Definitions}\label{sec:Halo size}
In order to compare the length scale $r_{\Lambda}$ to the size of a dark matter halo, we now turn to consider how the halo was formed in the first place. The simplest model is the spherical top-hat collapse model \cite{Liddle,Peebles,Peacock,White:2000jv}, where they consider the collapse of a spherically symmetric object of a uniform overdensity in a smooth background. By Birkhoff's theorem, the overdense region will evolve independently of the background density, which evolves as a spatially flat universe \cite{Liddle}. The Friedmann equation then describes the overdense sphere as a positively curved universe whose expansion rate initially matches to that of the background. Since the region is overdense, its expansion rate will slow down relative to that of the background, begin to recollapse, and eventually reach a singularity. In practice, however, the recollapse stops at virialization. We thus expect the final density of the overdense region to be some constant times the density of the background. That is,
\begin{equation}\label{eq:rho}
	\rho = \Delta\rho_c,
\end{equation}
where $\Delta$ is some constant and $\rho_c$ is the critical density. Detailed calculations of $\Delta$ can be found in \cite{Liddle,Peebles,Peacock}. Different halo mass definitions and the interpretations of $\Delta$ are reviewed in \cite{White:2000jv}. Simulations then find virialization turns out to occur at $\Delta \simeq 200$. 

Let us now compare this to our length scale (\ref{eq:r_lambda}). Since in the spherical top-hat collapse model the overdense region is uniformly distributed, the mass contained within a radius $r_\D$ is given by 
\begin{equation}
	M = \frac{4 \Delta}{3}\pi r^{3}_\Delta \rho_c,
\end{equation}
and we obtain the halo size
\begin{equation}\label{eq:r_delta}
	r_\Delta = \left(\frac{3M}{4\pi \rho_c \Delta}\right)^{1/3}.
\end{equation}
In order to compare our length scale $r_\Lambda$ to $r_\Delta$, we now express the latter in terms of the cosmological constant. Recall that we can write the critical density in terms of the Hubble constant $H_0$ as $\rho_c = 3H_0^2/8\pi G$, which is related to the cosmological constant as $\Lambda = \frac{3H_0^2}{c^2}\Omega_\Lambda$ in standard notation, giving
\begin{equation}\label{eq:halo_size}
	r_\Delta = \left(\frac{6GM}{\L c^2}\frac{\O_\L}{\D}\right)^{1/3}.
\end{equation}
This expression is of the same form as (\ref{eq:r_lambda}) up to a dimensionless constant dependent on $\O_\L$ and $\D$. Due to the cube root factor, the dimensionless constants $\O_\L$ and $\D$ have a diminished effect on this overall length scale. Substituting $\Delta = 200$ and $\Omega_\Lambda = 0.6911$ \cite{Ade:2015xua} into (\ref{eq:halo_size}), we have $r_\Lambda = 5.25~r_{200}$, that is they match to within a factor of order $1$.
%%%%%%%%%%%%%%%%%%%%%%%%%%%%%%%%%%%%%%%%%%%%%%%%%%%%%%%%%%%%%%%%%%%%%%%
\section{MOND and Verlinde's Emergent Gravity}\label{sec:MOND}
Finally, it is interesting to compare the geometric mean of length scales derived above with another characteristic length scale that arises in modified gravitational theories that try to describe galaxy rotation curves without dark matter, such as MOND \cite{Milgrom:1983ca}  and Verlinde's theory of emergent gravity \cite{Verlinde:2016toy}.

Milgrom \cite{Milgrom:1983ca} proposed that in the limit of small accelerations $a\ll a_0$, the acceleration of a test particle a distance $r$ from a mass $M$ is given by $a^2/a_0=GM/r^2$, where $a_0$ is a constant with dimensions of acceleration. For accelerations much larger than $a_0$, Newtonian gravity $a=GM/r^2$ is restored.
The two regimes may be interpolated to give 
\begin{equation}\label{eq:mu_MOND}
	\mu\left(\frac{a}{a_0}\right)a = \frac{GM}{r^2},
\end{equation}
where the function $\mu(x)$ satisfies $\mu(x) \approx 1$ when $x\gg1$, and $\mu(x) \approx x$ when $x\ll1$.  The acceleration constant $a_0$ defines the scale the Newtonian regime transitions to the MOND regime. The most recent and accurate numerical value of the acceleration constant was empirically determined to be about $a_0 \approx (1.2\pm0.2) \times 10^{-8}$ cm/s$^2$ \cite{Milgrom:2014usa}. Remarkably, this coincides within a factor of order $1$ with the cosmological acceleration scale $cH_0$. It is therefore interesting to convert this acceleration into a length scale on which new physics enters into galactic dynamics.

If we follow the simple de Sitter model of section \ref{sec:velocity calculation}, and evaluate (\ref{eq:mu_MOND}) at $a = a_0 = \g cH_0$ for some constant $\g$ of order $1$, and with $M$ identified as the mass enclosed, we get 
\begin{equation}\label{eq:r_MOND}
	r_{MOND} = \sqrt{\frac{GM}{c^2\g}\left(\frac{3}{\L}\right)^{\frac{1}{2}}}
\end{equation}
and we see the length scale is the geometric mean of the Schwarzschild radius and the cosmological horizon size. This is parametrically smaller than the other geometric mean found for the halo size (\ref{eq:halo_size}), though interestingly involves the same quantities.

Similar considerations apply to Verlinde's theory of emergent gravity \cite{Verlinde:2016toy}. Verlinde claimed that the observed phenomena in galaxies and clusters currently attributed to dark matter, such as the flattening of rotation curves, are the consequences of emergent gravity. In this theory, spacetime and gravity emerge from the entanglement structure of an underlying microscopic description. Consequently, there exists an extra dark gravitational force that can be effectively described by an apparent dark matter. Emergent gravity describes the amount of apparent dark matter in terms of the amount of baryonic matter for a spherically symmetric systems in non-dynamical situations.

According to emergent gravity, a strict area law for the entanglement entropy is required to derive the Einstein equations. Since de Sitter space has a cosmological horizon, it carries a finite entropy and temperature according to the Bekenstein-Hawking entropy and Hawking temperature respectively. From the horizon entropy and temperature, Verlinde proposed that microscopically, de Sitter space corresponds to a thermal state where part of its degrees of freedom are being thermalized. He then argued that due to thermalization there exists a thermal volume law contribution to the entanglement entropy. In particular, on galactic and larger scales, this volume law competes with the area law such that the microscopic de Sitter states do not thermalize and exhibit memory effects in the form of an entropy displacement caused by matter. This memory effect can be described effectively as an additional gravitational force from an apparent dark matter.

Balancing the entropic contribution of the apparent dark matter with the horizon entropy decrease due to the presence of baryonic matter leads to the relation \cite{Verlinde:2016toy}
\begin{equation}
	\frac{r}{L}\frac{A(r)}{4G\hbar} = \frac{2\p Mr}{\hbar}
\end{equation}
where $A(r)$ is the area enclosed at radius $r$, $M$ is the baryonic mass enclosed, and $L$ is the horizon size. For spherical symmetry, $A=4\p r^2$ and we are immediately led to the length scale
\begin{equation}
	r_{entropic} = \sqrt{2GML}
\end{equation}
which is exactly the geometric mean of the Schwarzschild radius and the cosmological horizon size $L$, matching (\ref{eq:r_MOND}) up to a factor of order $1$. Numerically this length scale is about $2$ orders of magnitude smaller than the halo size computed above, so clearly requires radically new physics to explain.
\section{Conclusion}
We have shown that there exists a length scale $r_\L$ which we interpret as a maximum size of a gravitationally bound system in the presence of a positive cosmological constant. The length scale is proportional to the geometric mean of the cosmological horizon size squared, and the Schwarzschild radius of the halo mass. This length scale is comparable to the dark matter halo size definition currently used in the astrophysics literature. We hope the very simple derivation of this scale offered here provides a helpful perspective on its physical meaning. We note the results here are somewhat at odds with some older work \cite{Cooperstock:1998ny}, though we certainly concur on scales well below the dark matter halo size.

The length scale at which new effects emerge was also computed for modified gravity theories, and found to have a similar geometric mean structure, though now being simply the geometric mean of cosmological size and Schwarzschild radius. This makes it manifest that cosmological effects can be ignored when considering galactic rotation curves in conventional gravity, and makes clear the length scale at which new physics will set in, in these new modified gravity approaches.

\section{Acknowledgments}
We thank T. Jacobson for pointing out reference \cite{Price:2005iv}. We also thank M. Mizony for pointing out references \cite{Souriau}, \cite{Mizony}, and \cite{Virgo}. This work is supported in part by DOE grant DE-SC0010010.

\end{document}